\documentclass[10pt, indentfirst]{jres-software}
\newcommand{\iu}{{i\mkern1mu}}

\definecolor{codegreen}{rgb}{0,0.6,0}
\definecolor{codegray}{rgb}{0.5,0.5,0.5}
\definecolor{codepurple}{rgb}{0.58,0,0.82}
\definecolor{backcolour}{rgb}{0.95,0.95,0.92}
\lstdefinestyle{mystyle}{
    commentstyle=\color{codegreen},
    keywordstyle=\color{magenta},
    numberstyle=\tiny\color{codegray},
    stringstyle=\color{codepurple},
    basicstyle={\footnotesize\ttfamily},
    breakatwhitespace=false,
    breaklines=true,
    captionpos=b,
    keepspaces=true,
    numbers=left,
    numbersep=5pt,
    showspaces=false,
    showstringspaces=false,
    showtabs=false,
    tabsize=4,
    numbers=none,
    frame = single,
    extendedchars=false,
    escapeinside={(*}{*)},
    language=Python,
    mathescape=true,
}
\usepackage[utf8]{inputenc}

\providecommand{\tightlist}{\setlength{\itemsep}{0pt}\setlength{\parskip}{0pt}}

\begin{document}

\title{pyLLE: a Fast and User Friendly Lugiato-Lefever Equation Solver}

\author[1,2,*]{Gregory Moille}
\author[1,2,3]{Qing Li}
\author[1,2]{Xiyuan Lu}
\author[1,4,$\dagger$]{Kartik Srinivasan}
\affil[1]{National Institute of Standards and Technology,Gaithersburg, MD 20899, USA}
\affil[2]{Institute for Research in Electronics and Applied Physics and Maryland Nanocenter, University of Maryland, College Park, Maryland 20742, USA}
\affil[3]{Electrical and Computer Engineering, Carnegie Mellon University, Pittsburgh, PA 15213, USA}
\affil[4]{Joint Quantum Institute, NIST/University of Maryland, College Park, Maryland 20742, USA}

\affil[*]{gregory.moille@nist.gov}
\affil[$\dagger$]{kartik.srinivasan@nist.giv}

\softwareDOI{2059}
\softwareVersion{2.1}
\acceptedDate{April 29, 2019}
\publisheddDate{To be added by Jres}
\keywords{Python, Julia, Micro-Combs, Non-Linear Optics}

\maketitle

\section{Introduction}
\label{sec:introduction}

The Lugiato-Lefever Equation (LLE), first developed to provide a description of spatial dissipative structures in optical systems~\cite{Lugiato:1987hu}, has recently made a significant impact in the integrated photonics community, where it has been adopted to help understand and predict Kerr-mediated nonlinear optical phenomena such as parametric frequency comb generation~\cite{DelHaye:2007gi,kippenberg_dissipative_2018} inside microresonators~\cite{Coen2012,Chembo2013,LLE_Gaeta_route}. The LLE is essentially an application of the nonlinear Schr\"odinger equation (NLSE)~\cite{Agrawal_NFO} to a damped, driven Kerr nonlinear resonator, so that a periodic boundary condition is applied. Importantly,  a slow-varying time envelope is stipulated, resulting in a mean-field solution in which the field does not vary within a round trip.  This constraint, which differentiates the LLE from the more general Ikeda map~\cite{Hansson:2015en}, significantly simplifies calculations while still providing excellent physical representation for a wide variety of systems.  In particular, simulations based on the LLE formalism have enabled modeling that quantitatively agrees with reported experimental results on microcomb generation~\cite{Weiner_soliton_Si3N4,Li2017} (e.g., in terms of spectral bandwidth), and have also been central to theoretical studies that have provided better insight into novel nonlinear dynamics that can be supported by Kerr nonlinear  microresonators~\cite{Kozyreff:2012fb,Luo2016a,Moille:2018ba}.

The great potential of microresonator frequency combs (microcombs) in a wide variety of applications~\cite{kippenberg_dissipative_2018,papp_microresonator_2014,suh_microresonator_2016,marin-palomo_microresonator-based_2017,spencer_optical-frequency_2018,trocha_ultrafast_2018,suh_soliton_2018,dutt_-chip_2018,obrzud_microphotonic_2019,suh_searching_2019} suggests the need for efficient and widely accessible computational tools to more rapidly further their development. Although LLE simulations are commonly performed by research groups working in the field, to our knowledge no free software package for solving this equation in an easy and fast way is currently available. Here, we introduce pyLLE, an open-source LLE solver for microcomb modeling.  It combines the user-friendliness of the Python programming language and the computational power of the Julia programming language.

\section{Software Specifications}
\label{sec:softwarespec}

\begin{table}[H]
    \centering
    \small
    \def\arraystretch{1.5}%  1 is the default, change whatever you need
    \begin{tabular}{|l|l|}
        \hline
        \textbf{NIST Operating Unit} & Physical Measurement Laboratory, Microsystems and Nanotechnology Division \\ \hline
        \textbf{Category} & Lugiato Lefever Equation Solver.  \\ \hline
        \textbf{Targeted Users} & Scientist / Researcher / Engineer in Photonics/Metrology \\ \hline
        \textbf{Operating Systems} & Windows/Linux/MacOS \\ \hline
        \textbf{Programming Language} & Python$>$3.4, Julia 0.6.4 \\ \hline
        \textbf{Inputs/Outputs} & Input text file with resonator dispersion; output electric field data. \\ \hline
        \textbf{Documentation} & \href{https://usnistgov.github.io/pyLLE/}{https://usnistgov.github.io/pyLLE/} \\ \hline
        \textbf{Accessibility} & N/A. \\ \hline
        \textbf{Disclaimer} & \href{https://www.nist.gov/director/licensing}{https://www.nist.gov/director/licensing} \\
        \hline
    \end{tabular}
\end{table}

\section{Lugiato-Lefever Equation Model}
\label{sec:model}

The Lugiato-Lefever equation is derived from the nonlinear Schr\"odinger equation, which can be written, if considering only the third order nonlinearity and in the reference frame moving with the group velocity, as:
\begin{equation}
    \frac{\partial E(z, \tau)}{\partial z} = - \frac{\alpha}{2}E + \iu \sum_{k>1} \frac{\beta_k}{k!}\left(\iu \frac{\partial}{\partial \tau} \right)^k E + \iu \gamma|E|^2 E
    \label{eq:NLSE}
\end{equation}

\noindent with $E(z,\tau)$ being the complex electric field propagating along $z$ during the time $\tau$, $\alpha$ the loss per unit length, $\gamma =  n_2\omega_0/(c_0 A_{\rm eff})$ the effective nonlinear coefficient, $n_2$ is the Kerr nonlinear coefficient (i.e., the nonlinear refractive index), $\omega_0$ the angular frequency of nearest mode close to the pump frequency $\omega_{pmp}$, $A_{\rm eff}$ the effective mode area in the resonator at the pump frequency, $\beta_{\rm k}$ is the $k$\textsuperscript{th}-order Taylor expansion coefficient of the dispersion~\cite{Agrawal_NFO}, and $c_0$ the speed of light in vacuum.

\begin{figure}[h]
    \begin{center}
        \includegraphics[width = \textwidth]{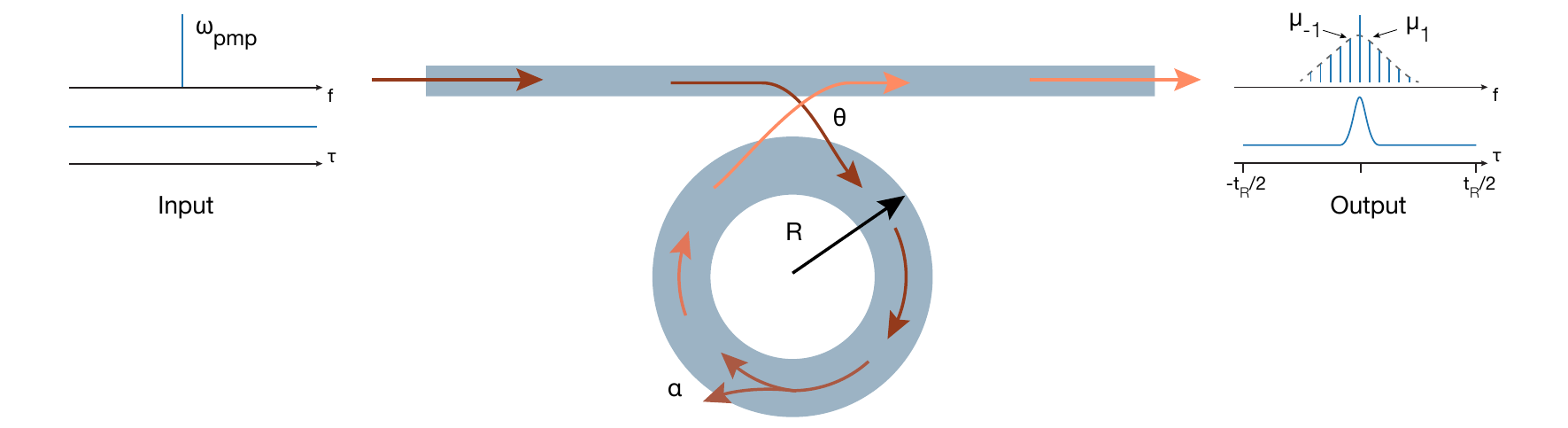}
        \caption{\label{fig:Schematic} Cartoon of a ring resonator coupled to a waveguide where a continuous wave pump is sent as an input and produces, through $\chi^{(3)}$-mediated nonlinear interactions in the ring, a frequency comb at the output. The coupling coefficient from the waveguide to the ring and vice-versa is $\theta$ and the intrinsic resonator losses per unite of length is $\alpha$.\vspace{-2ex}}
    \end{center}
\end{figure}

Next, we assume a geometry such as the one described in Figure~\ref{fig:Schematic}, where light injected into an optical waveguide is coupled into a traveling wave resonator (such as a ring or disk, for instance). Hence, a periodic condition for each round trip needs to be introduced, which leads to the coupled equations used to solve the Ikeda map~\cite{Hansson:2015en}:

\begin{eqnarray}
    &&E^{(m+1)}(0,\tau) = \theta \cdot E_{in} + (1-\theta) \cdot E^{(m)}\left( L,\tau \right)e^{i\phi_0}\label{eq:ikeda1}\\
    &&\frac{\partial E^{(m)}(z, \tau)}{\partial z} = - \frac{\alpha}{2}E^{(m)} + \iu \sum_{k>1} \frac{\beta_k}{k!}\left(\iu \frac{\partial}{\partial \tau} \right)^k E^{(m)} + \iu \gamma|E^{(m)}|^2 E^{(m)}
    \label{eq:ikeda2}
\end{eqnarray}

\noindent where $E^{(m)}$ represents the field in the resonator for the  $m$\textsuperscript{th} round trip, $\phi_0$ the phase shift accumulated during a single round trip, $L=2\pi R$ the length of the resonator and $\theta$ the resonator-waveguide coupling coefficient.

We then make a key assumption, that of  a slowly-varying field envelope, so that little non-linear physics happens during a single round trip. This assumption implies the possibility to average equations~(\ref{eq:ikeda1}) and (\ref{eq:ikeda2}) over one round trip, leading to the LLE:

\begin{equation}
    t_R\frac{\partial E(t, \tau)}{\partial t} = \left(- \left(\frac{\alpha'}{2} - \iu\delta_0\right) + \iu \sum_{k>1} \frac{\beta_k}{k!}\left(\iu \frac{\partial}{\partial \tau} \right)^k \right)E + \gamma|E|^2 E  + \sqrt{\theta}E_{in}
    \label{eq:LLE}
\end{equation}

\noindent with $t$ being the slow time and $\tau$ the fast time (i.e $-t_R/2\leq \tau \leq t_R/2$), $t_R$ the round trip time, $\alpha' = \alpha L  + \theta$ are the total round trip losses in the cavity and $\delta_0 = \omega_{pmp}-\omega{_0}$ the detuning of the pump field with respect to the closest cold-cavity resonance $\omega_0$.

For systems with a complex dispersion profile, such as microring resonators, instead of using a Taylor expansion of the dispersion that results in a set of different $\beta$ coefficients (i.e., dispersion to different orders), it is often more convenient and accurate to work with the integrated dispersion $D_{int} = \omega_\mu - \left(\omega_0 + D_1 \mu\right)$, with $\mu$ an integer representing the mode number relative to the pump resonance (for which $\mu$=0), $\omega_{\mu}$ is the angular frequency of the $\mu$\textsuperscript{th} mode, and $D_1$ is the angular free-spectral range evaluated at the pumped resonance (i.e., $D_1/2\pi$ is the free-spectral range in units of Hz). This implies that one has to work in the frequency domain. Considering the approximation of the slowly-varying envelope, one can consider the Fourier transform, $\mathrm{FT}$,  and inverse Fourier transform, $\mathrm{FT}^{-1}$, relative to the fast time $\tau$. 
Hence, equation~(\ref{eq:LLE}) becomes the one implemented in the package presented here:

\begin{equation}
    t_R \frac{\partial E(t, \tau)}{\partial t} = - \left(\frac{\alpha'}{2} - \iu\delta_0 \right)E + \iu \cdot \mathrm{FT}^{-1}\left[ -t_R D_{int}(\omega) \cdot \mathrm{FT}\left[E(t, \tau)\right]\right] + \gamma|E|^2 E  + \sqrt{\theta}E_{in}
    \label{eq:LLEdint}
\end{equation}

\noindent This approach is similar to the one found in Hansson and Wabnitz~\cite{Hansson:2015en}, for simulating the Kerr dynamics across a large number of modes using the Coupled Mode Theory (CMT) formalism. It is convenient indeed to work in the frequency domain, as it avoids the approximation of truncating the dispersion beyond a certain order, and the resonance frequencies can be found accurately using a Maxwell's equation eigenvalue solver. Moreover, working in the frequency domain allows the use of well-developed computational libraries for executing the Fourier transform operation in one iteration of the split-step Fourier technique, and is the approach implemented in the pyLLE package.

\section{pyLLE Description}

\subsection{Introduction to pyLLE}

Several approaches for solving the complex field of a multi-mode resonator have been studied, including Coupled Mode Theory (CMT) and the LLE. However, the LLE provides a temporal approach which is well suited to study Kerr non-linear resonators, especially in the context of soliton microcombs. Our fundamental goal in the pyLLE package is to provide a fast, efficient, platform-independent and freely available tool for simulating the LLE, to enable researchers with interest in microcombs to simulate their behavior without requiring in-depth knowledge of the LLE formalism or its numerical implementation.

The bulk of the computation is done through the Julia back-end. The Julia language provides an efficient computational interface with BLAS (basic linear algebra subprograms), enabling reuse of the same Fast Fourier Transform plan (i.e \textit{plan\_fft}) throughout several iterations of a loop, resulting in an overall simulation that lasts only a few minutes. Python provides complementary functionality, with easy scripting, display of figures, and saving of results.

To assess the performance of the solver, we perform the same simulation of a Si$_3$N$_4$ ring resonator (described in more detail in the documentation of ref.~\cite{pyLLEdoc}) using fixed hardware (a commercial laptop, the MacBook Pro 2017 3.1 GHz Intel Core i5, RAM 16 GB 2133 MHz LPDDR3) and three different software implementations, pyLLE, a pure Python version, and Matlab-based code~\cite{NIST_disclaimer}. We find that the simulation, in which the field behavior across 244 modes of the resonator is calculated, is an order of magnitude faster using pyLLE than the pure Python approach, and almost twice as fast as a Matlab solver (Table~\ref{tab1}). We note that Matlab has a proprietary license, compared to the open-source pyLLE package.

\begin{table}[htb]
 \centering \caption{\label{tab1}Performance comparison of the same resonator LLE simulation using three different software implementations.}
\begin{tabular}{ccc}
    \hline
    Matlab 2018a & Pure Python & \textbf{pyLLE} \\
    \hline
    19 mins & 45 min & 11 mins \\
    \hline
   \end{tabular}
\end{table}

As mentioned above, pyLLE is not only fast but highly user-friendly, thanks to the Python language which is becoming an increasingly valuable resource in the scientific toolbox. Below we describe a full example of how to use pyLLE, which consists ultimately of only a few lines of code. The complete example is accessible online~\cite{pyLLEdoc}.

\subsection{Example of pyLLE use}

In this section, we step through an example of how a simulation is set up and executed in pyLLE. We assume the pyLLE package has been installed on the user local machine following the instructions available in ref.~\cite{pyLLEdoc}.

\subsubsection{Setting up pyLLE}
\label{subsub:SetuppyLLE}

Let’s start by importing the package:

\begin{lstlisting}
import pyLLE
\end{lstlisting}

One has to define the resonator parameters.

\begin{lstlisting}[mathescape=true]
res =     {'R':  23e-6, # ring radius in meters
           'Qi': 1e6,   # Intrinsic Q factor
           'Qc': 1e6,   # Coupling Q factor
           'γ':  1.55,  # Effective nonlinear coefficient at the pump frequency                            (units of W$^{-1}$m$^{-1}$)
           'dispfile': 'TestDispersion.txt',
          }
\end{lstlisting}

The parameters in the code above are mostly self-explanatory and are chosen to fit the physical device investigated. The \textit{dispfile} parameter should be a raw text file, without header, in a ``comma separated value'' (csv) format consisting of the integer azimuthal mode orders and corresponding resonance frequencies (in Hz) as the first and second column, respectively. Importantly, the parameter names are the ones commonly used in the LLE formalism, hence some follow the Greek alphabet. To improve the usability of pyLLE, one can define a parameter either by its Greek symbol or through its Latin name (\textit{e.g} one can use either $\gamma$ or \textit{gamma}).

The next step is to define the simulation parameters through the dictionary:

\begin{lstlisting}
import numpy as np
sim = {'Pin':     150e-3,     # Input power in W
       'Tscan':   1e6,        # Simulation length in units of number of round trips
       'f_pmp':   191e12,     # Pump Frequency in Hz
       'δω_init': 2e9*2*np.pi,# Initial detuning of the pump in rad/s
       'δω_end': -8e9*2*np.pi,# End detuning of the pump in rad/s
       'μ_sim':  [-74,170],   # Azimuthal mode numbers to simulate in the LLE on the left and right side of the pump
       'μ_fit':  [-71,180],   # Azimuthal mode number range to fit the dispersion on the left and right side of the pump
       }
\end{lstlisting}

In the parameters described above, one specifies a simulation with pump power \textit{Pin} in the input coupling bus waveguide of 150~mW, with a linear detuning ramp of the pump from \textit{$\delta\omega$\_{init}} to \textit{$\delta\omega$\_{end}} relative to the pump mode angular frequency, \textit{i.e.} the mode closest to the defined pump frequency \textit{f\_{pmp}}. The simulation length \textit{Tscan} is in units of number of round trips, as is most convenient in the LLE formalism. Two parameter ranges for the simulation frequency range have to be defined, \textit{μ\_fit}, which determines the fit window for the raw data found in \textit{dispfile}, and \textit{μ\_sim} which is the number of modes simulated in the LLE. \textit{μ\_sim} can differ from \textit{μ\_fit} (i.e., the simulation can be performed over a truncated range or can be expanded over a larger range through extrapolation).

Once the resonator and simulation parameters are set, one can easily instantiate the pyLLE class:

\begin{lstlisting}
solver = pyLLE.LLEsolver(sim=sim,res=res)
\end{lstlisting}

\subsubsection{Dispersion Analysis}
\label{subsub:DispAnalysis}

To plot and retrieve all the dispersion data, the method \textit{solver.Analyze} has to be called, resulting in the plot of the integrated dispersion $D_{int}$, and returning the handle of the figure (and axes if using matplotlib). \\

\begin{minipage}[c]{0.3\textwidth}
\begin{lstlisting}
_ = solver.Analyze(plot=True, plottype='all')
\end{lstlisting}
\end{minipage}
\hfill
\begin{minipage}[c]{0.65\textwidth}
    \begin{center}
      \includegraphics[width = \textwidth]{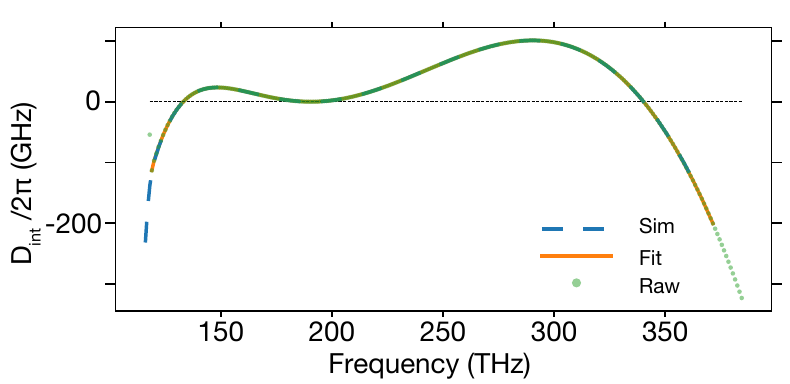}
      \captionof{figure}{\label{fig:Disp} Integrated dispersion retrieved through the \textit{dispfile} file, with the raw values from the file (green circles), fit dispersion (orange solid line), and integrated dispersion that will be used in the solver (blue dashed line).}
    \end{center}
\end{minipage}
\vspace{2ex}

Here, one extrapolates (orange curve - fit dispersion) outside of the spectral window containing the raw dispersion data (green circles), to enable a simulation over a broader spectral range (blue dashed line). One has to be careful about using this extrapolation feature, to make sure that the extrapolated dispersion is a reasonable physical representation of the dispersion.  For example, extrapolation can result in ripples in the integrated dispersion, creating zero-crossings that are artifacts that will influence the LLE simulation results.

\subsubsection{Temporal Simulation}

The core function of pyLLE is to solve the full temporal Lugiato-Lefever Equation as described in eq.~(\ref{eq:LLEdint}). The solver implemented in this package is based on a Julia core called from python. To interface both languages easily, a .hdf5 file is created in a temporary location with all the necessary data to solve the LLE.

We first set up the solver:
\begin{lstlisting}
solver.Setup()
\end{lstlisting}

After which we call the temporal solver :
\begin{lstlisting}
solver.SolveTemporal()
\end{lstlisting}

A progress bar has been implemented which displays the state of the current simulation, making it convenient to assess the time a simulation will take and when it is complete. To process the data through python, one has to retrieve the Julia data using the method:
\begin{lstlisting}
solver.RetrieveData()
\end{lstlisting}

\vspace{1ex}

The pyLLE class includes methods to ease the plotting process. The \textit{solver.PlotCombPower} method (fig.\ref{fig:CombPower}) provides an overview of the microcomb behavior, plotting the spectra, time envelope, and comb power inside the resonator for the different steps of the LLE, which are sub-sampled for faster display.\\
\begin{minipage}[c]{0.3\textwidth}
\begin{lstlisting}
solver.PlotCombPower()
\end{lstlisting}
\end{minipage}
\hfill
\begin{minipage}[c]{0.65\textwidth}
    \begin{center}
      \includegraphics[width = \textwidth]{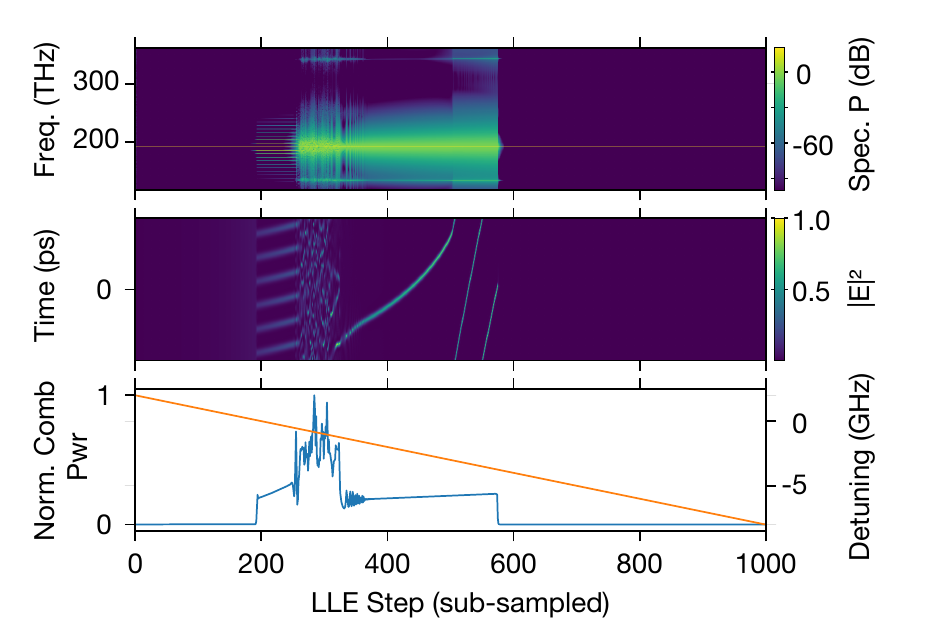}
      \captionof{figure}{\label{fig:CombPower} (a) Comb spectra vs. LLE step. (b) Temporal envelope of the electromagnetic field vs. LLE step. (c) Comb power (blue) vs. LLE step.  The pump detuning during the simulation (orange) is also plotted.}
    \end{center}
\end{minipage}
\vspace{2ex}

To obtain a better idea of the microcomb behavior for a given laser-cavity mode detuning, we can plot the spectrum and the temporal profile of the electric field in the resonator at a fixed LLE step:\\
\begin{minipage}[c]{0.3\textwidth}
\begin{lstlisting}
ind = 570
_ = solver.PlotCombSpectra(ind)
\end{lstlisting}
\end{minipage}
\hfill
\begin{minipage}[c]{0.65\textwidth}
    \begin{center}
      \includegraphics[width = \textwidth]{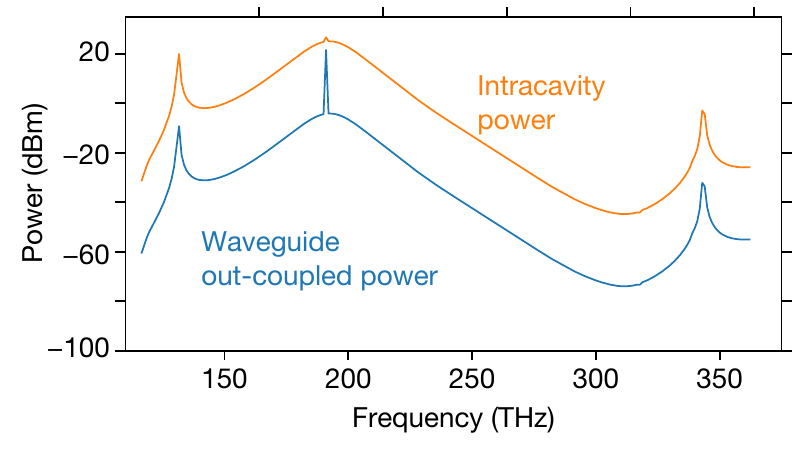}
      \captionof{figure}{\label{fig:CombSpectra} Spectrum of the soliton comb obtained at the LLE step = 570, inside the resonator (orange), and out-coupled into the access waveguide (blue).}
    \end{center}
\end{minipage}
\vspace{2ex}

The temporal profile can also be retrieved, which can be interesting, for example, in the case when we are on a soliton state for the given detuning, and by comparing the position of the soliton relative to the origin of time which is linked to the soliton drift:\\
\begin{minipage}[c]{0.3\textwidth}
\begin{lstlisting}
ind = 570
_ = solver.PlotSolitonTime(ind)
\end{lstlisting}
\end{minipage}
\hfill
\begin{minipage}[c]{0.65\textwidth}
    \begin{center}
      \includegraphics[width = \textwidth]{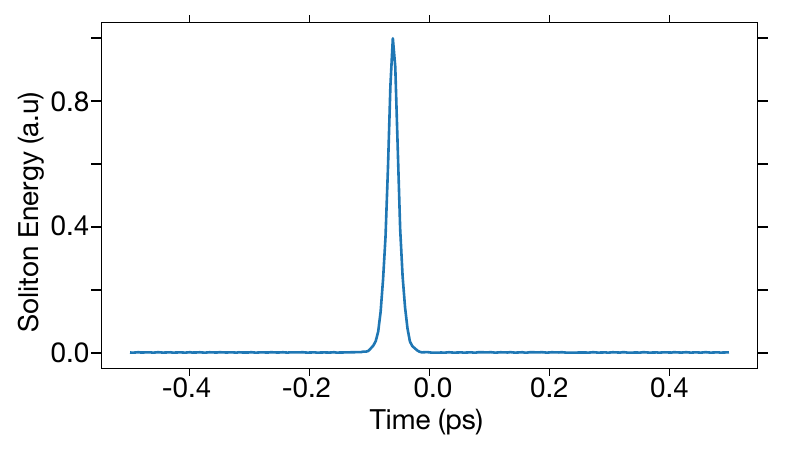}
      \captionof{figure}{\label{fig:FastTime} Electromagnetic envelope of the time-domain profile of a single soliton state inside the resonator. }
    \end{center}
\end{minipage}
\vspace{2ex}

\subsubsection{Steady-state Simulation}
Instead of solving the full temporal LLE in eq~(\ref{eq:LLEdint}), we can instead consider solutions to the steady-state LLE, i.e., considering the right-hand side of eq~(\ref{eq:LLEdint}) equal to zero. Steady-state solutions to the LLE indicate what types of states are supported by the system, but do not necessarily give an indication of how to access such states in practice.

Newton’s method can be used to find steady-state solutions and is implemented in the method \textit{solver.SolveSteadyState}. Although it gives fast results, the accuracy of such a solver remains questionable compared to a full temporal resolution of the equation (in particular, the convergence of the Newton’s method is not always assured).

To run the steady-state solver, one has to set it up in the same way as in the temporal method (see sections \ref{subsub:SetuppyLLE} to \ref{subsub:DispAnalysis} ), \textit{i.e.} defining the two dictionaries \textit{res} and \textit{sim} as in the example above, and calculating the dispersion profile. Here it is not necessary to use the method \textit{solver.Setup}, as the Newton's method places relatively low computational demands and thus everything is solved with Python.

However, the main difference with respect to the temporal solver is the introduction of a fixed detuning $\delta\omega$ of the pump relative to the closest mode, so that:

\begin{lstlisting}
solver.sim['δω'] = -5e9*2*np.pi # this value corresponds to the detuning at the end of the soliton step in the temporal simulation
\end{lstlisting}

The results of the steady-state Newton's method solver, with the function call indicated below, produces the spectrum shown in Fig.~\ref{fig:Steady} and returns the figure handle:\\

\begin{minipage}[c]{0.3\textwidth}
\begin{lstlisting}
steady_fig = solver.SolveSteadyState()
\end{lstlisting}
\end{minipage}
\hfill
\begin{minipage}[c]{0.65\textwidth}
    \begin{center}
      \includegraphics[width = \textwidth]{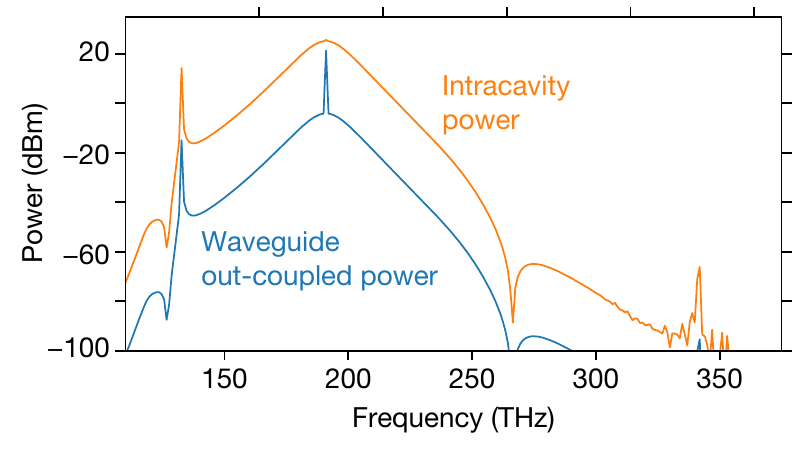}
      \captionof{figure}{\label{fig:Steady} Result of a steady-state simulation of the comb spectrum within the resonator (orange) and out-coupled into the access waveguide (blue).}
    \end{center}
\end{minipage}
\vspace{2ex}

% -------------------------------------
% \newpage
\section{pyLLE Module Description}

\subsection{\_llesolver.py}\label{llesolver.py}

\subsubsection{MyLogger(self, fname)}\label{myloggerself-fname}

Helper class for logging the results

\subsubsection{Latexify(self, **kwarg)}\label{latexifyself-kwarg}

Class that handles saving the figures in a nice way compatible with the column/page size of different latex templates.

\begin{itemize}
\tightlist
\item
  Input ({[}{]} = optional):

  \begin{itemize}
  \tightlist
  \item
    figname = name to save the figure (without extension)
  \item
    fig = matplotlib handle to the figure
  \item
    {[}fig\_width{]}: default = `1column'
  \item
    {[}frmt{]}: default = `pdf'
  \item
    {[}fig\_height{]} : default = 6.5
  \item
    {[}font\_size{]} : default = 8
  \end{itemize}
\end{itemize}

\subsubsection{LLEsolver(self, **kwargs)}\label{llesolverself-kwargs}

Class to solve the LLE Initialization input ({[}{]} = optional):

\begin{itemize}
\tightlist
\item
  res \textless{}dict\textgreater{}

  \begin{itemize}
  \tightlist
  \item
    Qi \textless{}float\textgreater{}: intrinsic Q of the resonator
  \item
    Qc \textless{}float\textgreater{}: coupling Q of the resonator
  \item
    R \textless{}float\textgreater{}: ring radius
  \item
    gamma \textless{}float\textgreater{}: Effective non-linearity of the
    waveguide comprising the resonator.
  \item
    dispfile \textless{}str\textgreater{} : str pointing to a .csv file
    where the azimuthal mode orders and corresponding resonance
    frequencies are saved
  \end{itemize}
\item
  sim \textless{}dict\textgreater{}

  \begin{itemize}
  \tightlist
  \item
    Tscan \textless{}float\textgreater{}: length of the simulation (in
    units of round trip)
  \item
    mu\_fit \textless{}list\textgreater{}: number of modes to fit
  \item
    mu\_sim \textless{}list\textgreater{}: number of mode that were
    simulated
  \item
    domega\_init \textless{}float\textgreater{}: initial detuning of the
    pump
  \item
    domega\_end \textless{}float\textgreater{}: final detuning of the
    pump
  \item
    {[}domga\_stop{]} \textless{}float\textgreater{}: where to stop the
    scan in detuning but keep doing the simulation
  \end{itemize}
\item
  debug \textless{}bool\textgreater{}: Save a trace in a log file in the
  working directory of the different actions pyLLE perform (default =
  True)
\end{itemize}

\paragraph{LLEsolver.Analyze(self, plot=False, f=None, ax=None, label=None, plottype=`all', zero\_lines=True, mu\_sim=None)}\label{llesolver.analyzeself-plotfalse-fnone-axnone-labelnone-plottypeall-zero_linestrue-mu_simnone}

\begin{itemize}
\item[]
Call pyLLE.analyzedisp.AnalyzeDisp to get the dispersion of the
resonator we want to simulate
\end{itemize}

\paragraph{LLEsolver.Setup(self)}\label{llesolver.setupself}
\begin{itemize}
\item[] Set up the simulation for the Julia back-end. Save the two main
dictionaries self.sim and self.res into a readable hdf5 file for Julia in
the temporary location defined by the operating system.
\end{itemize}

\paragraph{LLEsolver.SolveTemporal(self, tol=0.001, maxiter=6,
step\_factor=0.1)}\label{llesolver.solvetemporalself-tol0.001-maxiter6-step_factor0.1}
\begin{itemize}
\item[]
Call Julia to solve the LLE
\end{itemize}

\paragraph{LLEsolver.SolveSteadyState(self)}\label{llesolver.solvesteadysteateself}
\begin{itemize}
\item[]
Newton–Raphson method to find the root of the steady state equation
\end{itemize}

\paragraph{LLEsolver.RetrieveData(self)}\label{llesolver.retrievedataself}
\begin{itemize}
\item[]
Load the output hdf5 saved by Julia and transform it into a user-friendly
dictionary to be more suited for use with Python
\end{itemize}

\paragraph{LLEsolver.PlotCombPower(self,
do\_matplotlib=False)}\label{llesolver.plotcombpowerself-do_matplotlibfalse}
\begin{itemize}
\item[]
Plot a figure with 3 subplots.
\end{itemize}
\begin{itemize}
\item
  Top subplot = map of the spectra for the steps taken by the LLE (step
  sub-sampled to be 1000)
\item
  Middle subplot = temporal map of the intensity inside the resonator
  for the steps of the LLE
\item
  Bottom subplot = normalized comb power
\item
  Output
  \begin{itemize}
  \tightlist
  \item
    f \textless{}obj\textgreater{}: matplotlib/plotly figure handle
  \item
    ax \textless{}obj\textgreater{}: matplotlib axes handle (if plotly,
    only returns f)
  \end{itemize}
\end{itemize}

\paragraph{LLEsolver.PlotCombSpectra(self, ind,
f=None, ax=None, label=None, pwr=`both', do\_matplotlib=False,
plot=True)}\label{llesolver.plotcombspectraself-ind-fnone-axnone-labelnone-pwrboth-do_matplotlibfalse-plottrue}
\begin{itemize}
\item[]
Plot the spectrum for a given index in the 1000 sub-sampled LLE steps
\end{itemize}
\begin{itemize}
\tightlist
\item
  Input
  \begin{itemize}
  \tightlist
  \item
    ind \textless{}ind\textgreater{}: index in the LLE step for which the spectrum will be plotted
  \item
    f \textless{}obj\textgreater{}: matplotlib figure handle (if None,
    new figure)
  \item
    ax \textless{}obj\textgreater{}: matplotlib axes handle
  \item
    label \textless{}str\textgreater{}: label for the legend
  \item
    pwr \textless{}str\textgreater{}: `both', `ring', `wg' depending on
    the spectra wanted (inside the ring, the waveguide, or both)
  \end{itemize}
\item
  Output

  \begin{itemize}
  \tightlist
  \item
    freq \textless{}numpy.array\textgreater{}: frequency in Hz
  \item
    Sout \textless{}numpy.array\textgreater{}: spectral density of power
    in the waveguide (dBm)
  \item
    Sring \textless{}numpy.array\textgreater{}: spectral density of
    power in the ring (dBm)
  \item
    f \textless{}obj\textgreater{}: matplotlib/plotly figure handle
  \item
    ax \textless{}obj\textgreater{}: matplotlib axes handle (if plotly,
    only returns f)
  \end{itemize}
\end{itemize}

\paragraph{LLEsolver.PlotSolitonTime(self, ind, f=None, ax=None,
label=None,
do\_matplotlib=False)}\label{llesolver.plotsolitontimeself-ind-fnone-axnone-labelnone-do_matplotlibfalse}
\begin{itemize}
\item[]
Plot the spectra for a given index in the 1000 sub-sampled LLE step
\end{itemize}
\begin{itemize}
\item
  Input

  \begin{itemize}
  \tightlist
  \item
    ind \textless{}ind\textgreater{}: index in the LLE step to plot the
    spectra
  \item
    f \textless{}obj\textgreater{}: matplotlib figure handle (if None,
    new figure)
  \item
    ax \textless{}obj\textgreater{}: matplotlib axe handle
  \item
    label \textless{}str\textgreater{}: label for the legend
  \end{itemize}
\item
  Output

  \begin{itemize}
  \tightlist
  \item
    τ \textless{}obj\textgreater{}: Time in the resonator
  \item
    U \textless{}numpy.array\textgreater{}: Temporal electric field for
    the given step of the LLE
  \item
    f \textless{}obj\textgreater{}: matplotlib/plotly figure handle
  \item
    ax \textless{}obj\textgreater{}: matplotlib axes handle (if plotly,
    only returns f)
  \end{itemize}
\end{itemize}

\paragraph{LLEsolver.SaveResults(self, fname,
path=`./')}\label{llesolver.saveresultsself-fname-path.}

Save the whole class using the \textit{pickle} package, which makes it easy to load back the saved results

\begin{itemize}
\tightlist
\item
  Input

  \begin{itemize}
  \tightlist
  \item
    fname \textless{}str\textgreater{}: name to save. The `.pkl'
    extension will be added
  \item
    path \textless{}str\textgreater{}: path to save the results
    (defaults `./')
  \end{itemize}
\end{itemize}

\subsection{\_analyzedisp}\label{analyzedisp}

\subsubsection{AnalyzeDisp(self, **kwargs)}\label{analyzedispself-kwargs}

Call to analyze the dispersion of a simulated resonator. Initialization
input. Everything is in SI ({[}{]}=optional):

\begin{itemize}
\tightlist
\item
  Input:

  \begin{itemize}
  \tightlist
  \item
    f\_center \textless{}float\textgreater{}: pump frequency
  \item
    file \textless{}str\textgreater{}: .txt file to load
  \item
    R \textless{}float\textgreater{}: radius of the resonator
  \item
    rM\_fit \textless{}list\textgreater{}.: lower and upper bounds of
    mode to fit the dispersion
  \item
    rM\_sim \textless{}list\textgreater{}.: lower and upper bounds of the
    modes to extrapolate for the simulation
  \item
    f \textless{}obj\textgreater{}: matplotlib figure handle
  \item
    ax \textless{}obj\textgreater{}: matplotlib axes handle
  \item
    label \textless{}list\textgreater{}: list of the string for each
    plot to be labeled
  \item
    plottype \textless{}str\textgreater{}: define the type of plot `all'
    {[}defaults{]}, `sim', `fit', `fem', `ind'
  \end{itemize}
\end{itemize}

\paragraph{AnalyzeDisp.GetDint(self)}\label{analyzedisp.getdintself}

Retrieve the dispersion of a resonator based on the frequency of
resonance and azimuthal mode order. The data are fit using a cubic
spline method

\begin{itemize}
\tightlist
\item
  Output:

  \begin{itemize}
  \tightlist
  \item
    self.PrM\_fit: scipy.interpolate object which fitted the data
  \item
    self.Dint\_fit: fitted integrated dispersion for the simulated mode
  \item
    self.neff\_pmp: effective index at the pump frequency
  \item
    self.ng\_pmp: group index at the pump frequency
  \end{itemize}
\end{itemize}

\section{Acknowledgments}

\noindent The authors acknowledge funding support from the DARPA DODOS and ACES programs, the NIST-on-a-chip program, and the Cooperative Research Agreement between the University of Maryland and NIST-CNST (award no. 70NANB10H193). They also thank Dr. Su-Peng Yu and Dr. Jared H. Strait for their helpful comments on the manuscript.\\

\renewcommand{\bibname}{References}

% -------------------------------------
{
\footnotesize
\bibliographystyle{jresnist}
\bibliography{MoilleLLEsoftware}}

\vspace{20pt}

\noindent\textit{\textbf{About the authors:} Gregory Moille, Qing Li, and Xiyuan Lu are guest research associates in the Microsystems and Nanotechnology Division at NIST, and are postdoctoral scholars at the University of Maryland. Kartik Srinivasan is a physicist in the Microsystems and Nanotechnology Division at NIST and an adjunct professor at the University of Maryland.
The National Institute of Standards and Technology is an agency of the U.S. Department of Commerce.} \\
%% The last sentence here is required

\end{document}